\documentclass{article}

\makeatletter
\newdimen\mathindent
\def\equation{\@beginparpenalty\predisplaypenalty
  \@endparpenalty\postdisplaypenalty
  \refstepcounter{equation}\trivlist \item[]\leavevmode
  \hbox to\linewidth\bgroup $\m@th\displaystyle\hskip\mathindent}

\def\endequation{$\hfil\displaywidth\linewidth\egroup\hskip-12pt\@eqnnum\endtrivlist}
\makeatother

\begin{document}

\setcounter{page}{1}

\LRH{O. Boussa\"{i}d, J. Darmont,  F. Bentayeb and S. Loudcher}

\RRH{Warehousing complex data from the Web}

\VOL{x}

\ISSUE{x}

\BottomCatch

\PAGES{xxxx}

\CLline

\PUBYEAR{2006}

\subtitle{}

\title{Warehousing complex data from the Web}

\authorA{O. Boussa\"{i}d$^{*}$,  J. Darmont$^{*}$,     \newline F. Bentayeb and S. Loudcher}

\affA{ERIC, University of Lyon 2\\
5 avenue Pierre Mend\`{e}s-France\\
69676 Bron Cedex\\
France\\
Fax: +33 478 772 375\qquad E-mail: firstname.lastname@univ-lyon2.fr
\newline $^{*}$Corresponding authors }

\begin{abstract}
The data warehousing and OLAP technologies are now moving onto handling complex data that mostly
originate from the Web. However, intagrating such data into a decision-support process requires
their representation under a form processable by OLAP and/or data mining techniques.

We present in this paper a complex data warehousing methodology that exploits XML as a pivot
language. Our approach includes the integration of complex data in an ODS, under the form of
XML documents; their dimensional modeling and storage in an XML data warehouse; and their
analysis with combined OLAP and data mining techniques. We also address the crucial issue of
performance in XML warehouses.
\end{abstract}

\KEYWORD{Data warehousing, Web data, Complex data, ETL process, Dimensional modeling,
XML warehousing, OLAP, Data mining, Performance.}

\REF{to this paper should be made as follows: Boussa\"{i}d, O.,
Darmont, J., Bentayeb, F. and Loudcher, S.
(xxxx) 
`Warehousing complex data from the Web',
{\it Int. J. Web Engineering and Technology},
Vol. x, No. x, pp.xxx\textendash xxx.}

\BIO{Omar Boussa\"{i}d is an associate professor in computer science at the School of Economics
and Management of the University of Lyon 2, France. He received his PhD degree in computer
science from the University of Lyon 1, France in 1988. Since 1995, he has been in charge of
the Master's degree ``Computer Science Engineering for Decision and Economic Evaluation" at
the University of Lyon 2. He is a member of the Decision Support Databases research group within
the ERIC laboratory. His main research subjects are data warehousing, multi-dimensional databases
and OLAP. His current research concerns complex data warehousing, XML warehousing, data mining-based
multidimensional modeling, OLAP and data mining combining, and mining metadata in RDF form.\\[10pt]
J\'{e}r\^{o}me Darmont received his PhD degree in computer science from the University
of Clermont-Ferrand II, France in 1999. He has been an associate professor at the University of Lyon 2,
France since then, and became the head of the Decision Support Databases research group within
the ERIC laboratory in 2000. His current research interests mainly relate to the evaluation and
optimization of database management systems and data warehouses (benchmarking, auto-administration,
optimization techniques...), but also include XML and complex data warehousing and mining, and medical
or health-related applications.\\[10pt]
Fadila Bentayeb has been an associate professor at the University of Lyon 2, France since 2001.
She is a member of the Decision Support Databases research group within the ERIC laboratory.
She received her PhD degree in computer science from the University of Orl\'{e}ans, France in 1998.
Her current research interests regard database management systems, including the integration of
data mining techniques into DBMSs and data warehouse design, with a special interest for schema
evolution, XML and complex data warehousing, benchmarking and optimization techniques.\\[10pt]
Sabine Loudcher Rabaseda is an associate professor in computer science at the Department of
Statistics and Computer Science of the University of Lyon 2, France. She received her PhD degree
in computer science from the University of Lyon 1, France in 1996. Since 2000, she has been a
member of the Decision Support Databases research group within the ERIC laboratory. Her main
research subjects are data mining, multi-dimensional databases, OLAP, and complex data. Since 2003,
she has been the assistant director of the ERIC laboratory.}

\maketitle


\section{Introduction}

Decision-support technologies, including data warehouses and OLAP (On-Line Analytical Processing),
are nowadays technologically mature. However, their complexity makes them unattractive to many
companies; hence, some vendors develop simple Web-based interfaces [\ref{lawton06}].
Furthermore, many decision-support applications necessitate external data sources.
For instance, performing competitive monitoring for a given company requires the
analysis of data available only from its competitors. In this context, the Web is
a tremendous source of data, and may be considered as a farming system [\ref{hackathorn00}].

There is indeed a clear trend toward on-line data warehousing, which gives way to
new approaches such as virtual warehousing [\ref{belanger99}] or XML warehousing
[\ref{baril03}, \ref{hummer03}, \ref{nassis05}, \ref{park05}, \ref{pokorny02},
\ref{vrdoljak05}, \ref{zhang05}]. However, data from the Web are not only numerical
or symbolic, but may be:

\begin{BL}
\item   represented in various formats (databases, texts, images, sounds, videos...);
\item   diversely structured (relational databases, XML documents...);
\item   originating from several different sources;
\item   described through several channels or points of view (a video and a text that describe
the same meteorological phenomenon, data expressed in different scales or languages...);
\item   changing in terms of definition or value (temporal databases, periodical surveys...).
\end{BL}
We term data that fall in several of the above categories complex data [\ref{darmont05}].
 Managing such data involves a lot of different issues regarding their structure, storage
 and processing [\ref{darmont06}]; and classical data warehouse architectures must be
 reconsidered to handle them.

\begin{figure*}[hbt]
\caption{Complex data warehousing and analysis process}
\label{process} \vspace{-5cm}\epsfxsize=8cm \epsffile[0 0 500
600]{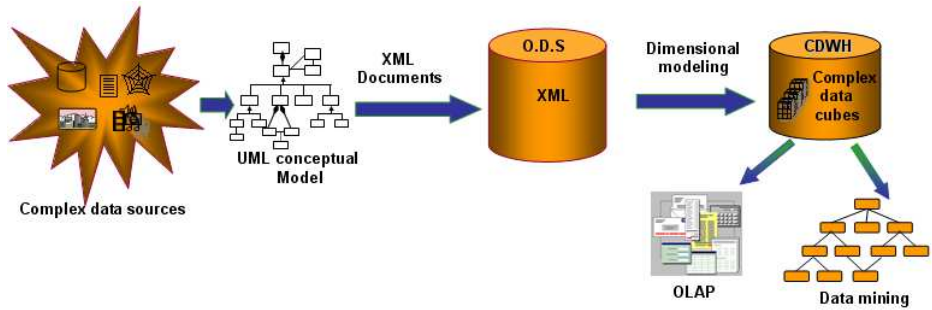}
\end{figure*}

In this context, our motivation is to integrate complex data from
the Web into a decision-support process, which requires the
integration and the representation of complex data under a form that
can be processed by on-line analysis and/or data mining techniques
[\ref{darmont03}]. We propose a full, generic data warehousing and
on-line analysis process (Figure~\ref{process}) that includes two
broad axes:
\begin{BL}
\item data warehousing, including complex data integration and modeling;
\item complex data analysis.
\end{BL}

To preserve the generic aspect of this process, we must exploit a
universal formalism for modeling and storing any form of complex
data. The XML (eXtensible Markup Language) formalism has emerged as
a dominant W3C~\footnote{http://www/w3.org} standard for describing
and exchanging semi-structured data among heterogeneous data
sources. Its self-describing hierarchical structure enables a
manipulative power to accommodate complex, disconnected, and
heterogeneous data. Furthermore, XML documents may be validated
against an XML Schema. It allows to describe the structure of a
document and to constraint its contents. With its vocation for
semi-structured data exchange, the XML language offers a great
flexibility for representing heterogeneous data, and great
possibilities for structuring, modeling, and storing them.

The approach we propose consists in representing complex data as XML
documents and in physically integrating them into an Operational
Data Storage (ODS), which is a buffer ahead of the actual data
warehouse. Then, we recommend an additional layer to model complex
data and prepare them for analysis. Complex data under the form of
XML documents are thus multidimensionally modeled to obtain an XML
data warehouse. Finally, complex data analysis can take place from
this warehouse, with on-line analysis, data mining or a combination
of two approaches.

One originality of our approach is that we do not envisage data mining only as a front-end,
stand-alone analysis tool. We indeed exploit data mining techniques throughout the whole
complex data warehousing process:

\begin{BL}
\item at the data integration stage, e.g., to extract semantic information from complex data;
\item in the multidimensional modeling phase, e.g., to discover pertinent measures or dimensions;
\item when analyzing data, e.g., by coupling OLAP and data mining;
\item in support to database administration, e.g., to automatically select pertinent indexes or materialized views.
\end{BL}

The objective of this paper is to present the issues we identified on the path
to designing and implementing this approach, as well as the solutions we devised
to solve them. The remainder of this paper is organized as follows.
Section~\ref{sec:ComplexDataIntegration} presents our approach for complex data
integration. Section~\ref{sec:XWarehousing} details the \emph{X-Warehousing} XML warehousing platform.
Section~\ref{sec:ComplexDataAnalysis} describes sample complex data analyses.
Section~\ref{sec:PerformanceIssues} addresses performance problems in XML data warehouses.
Finally, Section~\ref{sec:Conclusion} concludes this paper and discusses open issues.


\section{Complex data integration}
\label{sec:ComplexDataIntegration}

\subsection{Context and issues}

Companies collect    huge amount of heterogeneous  and complex data.
They aim at integrating these data in their Decision Support Systems
(DSS), and some efforts are needed  to structure them and to make
them as homogeneous as possible. In data warehousing, the prime
objective of storing data is to facilitate the decision process. To
achieve the value of a data warehouse, incoming data must be
transformed into an analysis-ready format. In the case of numerical
data, data warehousing systems often provide tools to assist in this
process. Unfortunately, standard tools are inadequate for producing
relevant analysis axis when data are complex. In such cases, the
data warehousing process should be adapted in response to evolving
data and information requirements. We need to develop tools to
provide the needed  analysis.

In a data warehousing process, the data integration phase is
crucial. Data integration is a hard task that involves
reconciliation at various levels (data models, data schema, data
instances, semantics). Indeed, the special nature of complex data
poses different and new requirements to data warehousing
technologies, over those posed by conventional data warehouse
applications. Hence, to integrate complex data sources, we need
more than a tool for organizing data into a common syntax.

Two main and opposed approaches are used to perform data
integration over heterogeneous data sources. In the mediator-based
approach [\ref{rousset02}], the different data remain located at their
original sources. User queries are executed through a
mediator-wrapper system [\ref{goasdoue00}]. A mediator reformulates queries
according to the content of the various accessible data sources,
while the wrapper extracts the selected data from the target
source. The major advantage of this approach is its flexibility,
since mediators are able to reformulate and/or approximate queries
to better satisfy the user. However, when the data sources are
updated, modified data are lost, which is not pertinent in a
decision-support context where historicity of data is important.

On the opposite, in the data warehouse approach [\ref{inmon96}, \ref{kimball96}], all
the data from the various data sources are centralized in a new
multidimensional database, the data warehouse. In a data warehouse
context, data integration corresponds to the ETL (Extract,
Transform, Load) process that accesses to, cleans  and transforms
the heterogeneous data before they are loaded into the data
warehouse. This approach supports the dating of data and is
tailored for analysis.

In this section, we present our approach for complex data
integration based on both data warehouse technology and multi-agent
systems (MAS). Our aim is to take advantage of the MAS, which are
intelligent programs composed of a set of agents, each one offering
a set of services, to achieve complex data integration. Indeed, we
can assimilate the three steps of the ETL process to services
carried out by specific agents.

\subsection{Proposed solution}

\emph{Complex data ETL process.}
The classical ETL approach proceeds in three steps. The first {\it
extraction} phase includes understanding and reading the data
source, and copying the necessary data in a buffer called the
preparation zone. Then, the second {\it transformation} phase
proceeds in successive steps: clean the data from the preparation
zone; discard some useless data fields; combine the data sources;
and build aggregates to optimize the most frequent queries. In
this phase, metadata are essential to store the transformation
rules and various correspondences. The third {\it loading} phase
stores the prepared data into multidimensional structures (data
warehouse or data marts).

To achieve complex data integration following the warehouse
approach, the traditional ETL process is ill-adapted. We present
here our approach to accomplish the ETL process in an original way.
We propose a modeling process to achieve the integration of complex
data into a data warehouse [\ref{boussaid03Holomas}]. We first
design a conceptual UML model for a complex object
[\ref{boussaid06jgo}]. The UML conceptual model is then directly
translated into an XML Schema, which we view as a logical model. The
obtained logical model may be either mapped into a relational,
object-relational or XML-native database.

\emph{Complex data UML model.} First, we present a generic UML model
that allows us to model not only low-level but also semantic
information concerning the complex data to be analyzed. After we
transform this UML model into an XML grammar (as a DTD -Document Type Definition- or an XML
Schema),  we generate the XML documents describing the complex data.
Therefore, we integrate complex data as XML documents into an ODS as
a first step in complex data warehousing.

\begin{figure*}[hbt] 
\caption{Generic complex data UML model}
\epsfxsize=9cm \centerline{\epsffile[0.55 0.55 390 495]{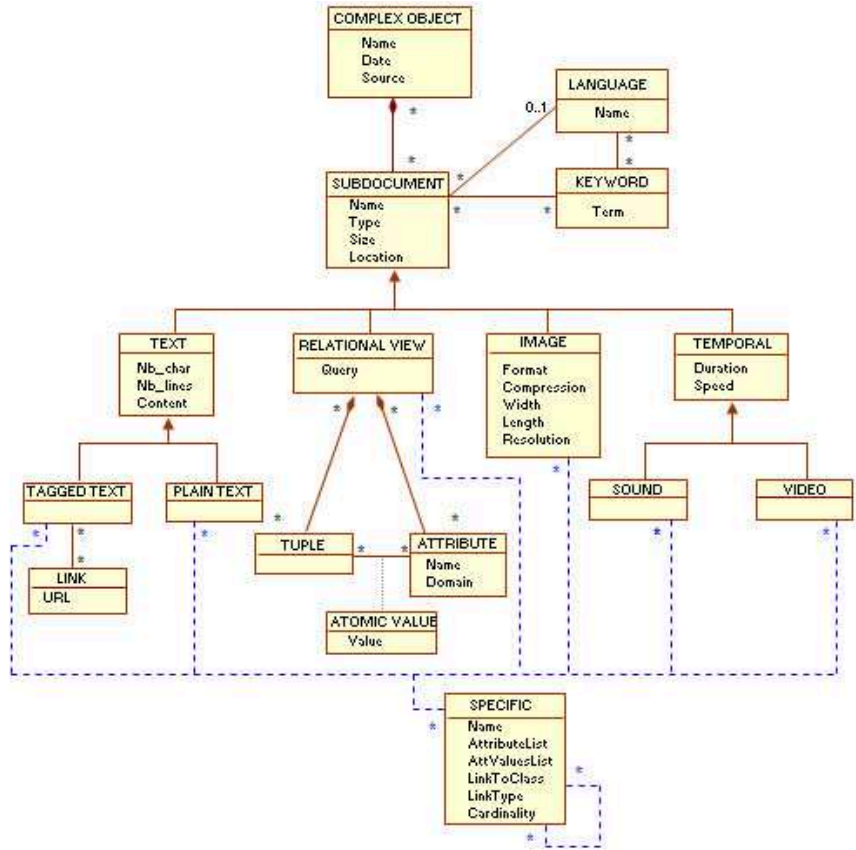}}
\label{myuml}
\end{figure*}

We choose to integrate the characteristics of data rather than the
original data themselves. We use XML  to describe the
characteristics of our complex data as it encloses not only the
content of the complex data, but also the way they are structured.
The basic characteristics (e.g., \emph{file size}, \emph{file name},
\emph{duration} for films or sounds; \emph{resolution} for images,
and so on) can be extracted automatically. These characteristics
capture low-level information concerning the original data. The
generic model that we present (Figure \ref{myuml}) allows us to add
semantic characteristics of data in order to enrich their
description by manual annotations or by knowledge automatically
extracted from data mining techniques.

The UML class diagram represents a complex object generalizing all
complex data types. Note that our goal here is to propose a
general data structure: the list of attributes for each class in
this diagram is willingly not exhaustive. Our generic model
defines a complex object which is composed of complex data
represented as subdocuments. The subdocuments have special
predefined low-level characteristics that depend on the type of
complex data they  contain. The (meta)class \emph{Specific} in
our model is  a generic class that allows  to define new
classes and relationships in the UML model, and thus enables
modeling semantic characteristics of complex data. It allows
not only to describe the semantic properties of data, but also any
other useful characteristic. Every class linked to the class
\emph{Specific} in the UML model can take advantage from it, when
instantiating the model (at implementation time), by defining its own
new characteristics.

\emph{MAS-based ETL approach.} A MAS is  a collection of actors that
communicate with each other [\ref{sycara96}]. Morover, each agent
(actor) is able to offer specific services and has a well-defined
goal. Each agent is able to perform several tasks, in an autonomous
way, and communicates the results to a receiving actor (human or
software). A MAS must respect the programming standards defined by
the Foundation for Intelligent Physical Agents (FIPA)
[\ref{fipa02}].

\begin{figure*}[hbt] 
\caption{MAS-based ETL architecture for complex data integration}
\epsfxsize=10cm \centerline{\epsffile{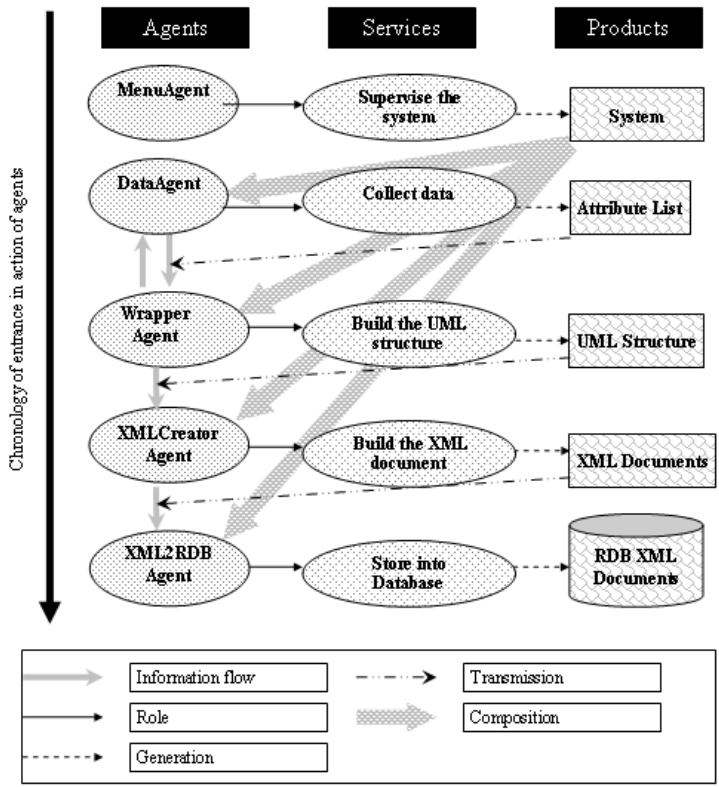}}
\label{system}
\end{figure*}

Our MAS-based integration approach is a flexible and evolutive
architecture on which we can add, remove or modify services, and
even create new agents (Figure \ref{system}) [\ref{boussaid03}]. To validate our
approach, we have developed a MAS-based ETL prototype: SMAIDoC,  which is
freely available on-line [\ref{bdd03}].

We have instantiated five agents that allow the integration of
complex data. The purpose of this collection of agents is to perform
several tasks. The first main agent in our prototype,
\emph{MenuAgent}, pilots the system, supervises agent migrations,
and indexes the accessible sites from the platform. Some others
default pilot agents help in the management of the agents and
provide an interface for the agent development platform. Two agents
named \emph{DataAgent} and \emph{WrapperAgent}, respectively, model
the input complex data into UML classes. Finally, the
\emph{XMLCreator} agent translates UML classes into XML documents
that are mapped into a relational database by the
\emph{XML2RDBAgent} agent  or that stored as a collection of XML
documents.

To develop our prototype, we have built a platform  using JADE
version 2.61 [\ref{jade02}] and the Java language [\ref{sun02}], which is
portable across agent programming platforms.

\subsection {Perspectives}

From a technical point of view, we can extend the services offered
by SMAIDoC, especially for extracting data from their sources and
analyzing  them. For example, the \emph{DataAgent} agent could
converse with online search engines and exploit their answers. We
could also create new agents in charge of modeling data in the
multidimensional way, and applying analysis methods such as OLAP or
data mining.

Finally, we aim at studying a meta-data representation of the
results of data mining techniques that generate rules, in mixed
structures combining XML-Schema and the Resource Description
Framework (RDF). These description languages are indeed well-suited
for expressing semantic properties and relationships between
meta-data.

Therefore, SMAIDoC is designed as an incremental and progressive
platform and as a technical support for our complex data integration
method, whose main objective is to describe and store complex data
into the XML documents.


\section{The X-Warehousing platform}
\label{sec:XWarehousing}

\subsection{Context and issues}

In our approach for complex data integration, we chose XML to
describe and to store data in an ODS. At this stage, it is possible
to mine the stored XML documents directly with the help of adapted
techniques such as XML structure mining, for instance (Section
\ref{xmlmining}). Otherwise, in order to analyse these XML documents
efficiently, it is interesting to warehouse them. Therefore, new
efforts are needed to integrate XML in classical business
applications. Feeding data warehouses with XML documents is also
becoming a challenging issue, since the multidimensional
organization of data is quite different from the semi-structured
organization. The difficulty consists in carrying out a
multidimensional design within a semi-structured formalism like XML.

Nevertheless, in the literature, we distinguish two separate
approaches in this field. The first approach focuses on the physical
storage of XML documents in data warehouses. XML is considered an
efficient technology to support data within structures well-suited
for interoperability and information exchange, and can definitely
help in feeding data warehouses. Baril and Bellahs\`{e}ne introduce
the View Model, onto which they build an XML warehouse called DAWAX
(DAta WArehouse for XML) [\ref{baril03}]. H\"{u}mmer et al. propose
an approach that focuses on the exchange and the transportation of
data cubes over networks, rather than multidimensional modeling with
XML [\ref{hummer03}].

The second approach aims at using XML to design data warehouses
according to classical multidimensional models such as star and
snowflake schemas. Pokorny uses a sequence of DTDs to explicit
dimension hierarchies that are logically associated, about the same
way referential integrity is achieved in relational databases
[\ref{pokorny02}]. Golfarelli et al. introduce a Dimensional Fact
Model represented by Attribute Trees [\ref{golfarelli99}]. They also
use XML Schemas to express multidimensional models, by including
relationships in subelements. Trujillo et al. also provide a DTD
model from which valid XML documents are generated to represent
multidimensional models at a conceptual level [\ref{trujillo04}].
Nassis et al. propose a similar approach, where an Object Oriented
(OO) standard model is used to develop a conceptual model for XML
Document Warehouses (XDW) [\ref{nassis04}]. An XML repository,
called xFACT, is built by integrating OO concepts with XML Schemas.
They also define virtual dimensions by using XML and UML package
diagrams in order to help the construction of hierarchical
conceptual views.

Since we are able to describe complex data in XML documents and we
need to prepare them to future OLAP analysis, storing them in a data
repository is not a sufficient solution. We rather need to express
through these documents a more interesting abstraction level that is
completely oriented toward analysis objectives. To achieve this
goal, we propose an approach, called X-Warehousing
[\ref{boussaid06}], which is entirely based on XML, to warehouse
complex data. It allows to build a collection of homogeneous XML
documents. Each document corresponds to an OLAP fact where the XML
formalism structures data according to a multidimensional model.

\subsection{Proposed solution}

\begin{figure*} [hbt] \caption{Overview of the \emph{X-Warehousing} approach}
\label{our_approach} \vspace{-6,5cm} \epsfxsize=10cm \centerline{\epsffile[0 0
250 350]{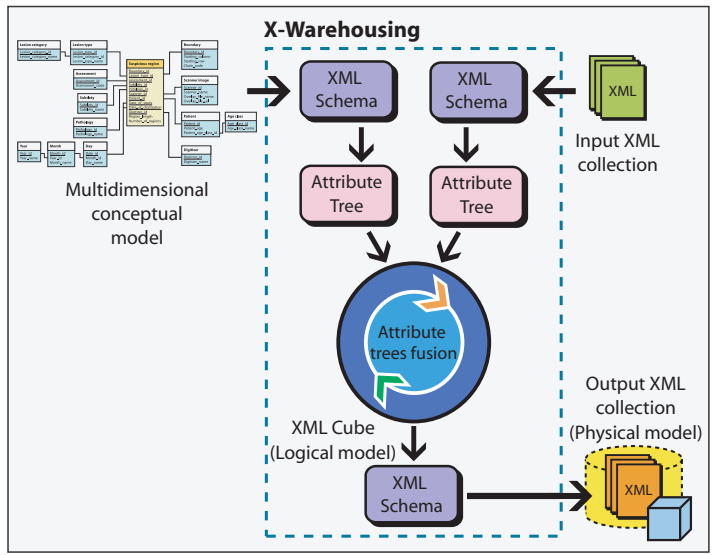} } \end{figure*}

We include in our approach a methodology that enables the use of XML as a
logical modeling formalism for data warehouses. This methodology starts from
analysis objectives defined by users according to a multidimensional conceptual
model (MCM). Therefore,  we focus on analysis needs rather than on the data
themselves.  The \emph{X-Warehousing} approach (Figure \ref{our_approach})
accepts a reference MCM and XML documents in
input. In fact, through the reference MCM, a user may design a data warehouse by
defining facts, dimensions, and hierarchies.  Despite the use of a star
schema or snowflake schema, the MCM depicts an analysis context
independently from its logical and physical representation. The MCM is then
transformed into a logical model via an \emph{XML Schema} (XSD file).

\vspace{0,1cm} In a second step, an \emph{attribute tree} is automatically generated from this
XSD file. An \emph{attribute tree} is a directed, acyclic and
weakly connected graph that represents a warehouse schema [\ref{golfarelli01}].
Once the reference  model is defined, we can submit XML documents to feed the designed warehouse.
\emph{XML Schemas} and \emph{attribute trees} are also extracted from the input XML
documents. We transform the reference model and the XML documents into
\emph{attribute trees} in order to make them comparable. In fact, two
\emph{attribute trees} can easily be merged together through a fusion process based on
\emph{pruning} and \emph{grafting} functions~[\ref{golfarelli01}]. At this
stage, two cases are possible: (1) if an input document contains the minimum
information required in the reference MCM, the document is accepted and merged
with the MCM. An instance of the XML documents is created and validated against
the resulted \emph{XML Schema}. This new \emph{XML Schema} represents the
logical model of the final \emph{XML Cube}; (2) if a submitted document
does not contain enough information to represent an OLAP fact according to the
reference MCM, the document is rejected and no output is provided. The
goal of this condition is to obtain an homogeneous collection of data with
minimum information to feed the final \emph{XML Cube}.

The interest of our approach is quite important since organizations
are treating domains of complex applications. In these applications,
a  special consideration is given to the integration of heterogenous
and complex information in DSS. For example, in \emph{breast cancer
researches} [\ref{heath00}], experts require efficient
representations of mammographic exams. Note that information about a
mammogram comes from different sources like texts,  annotations by
experts, and radio scanners. We think that structuring such a set of
heterogenous data within an XML format is an interesting solution
for warehousing them. Nevertheless, this solution is not sufficient
for driving future analyse. We propose to structure these data in
XML with respect to the multidimensional reference model of a data
warehouse. Output XML documents of the \emph{X-Warehousing} process
represent the physical model of the data warehouse. Each output
document corresponds to the multidimensional structured information
of an OLAP fact.

\emph{Modeling a warehouse with XML}. According to the proprieties
of the XML  documents, we propose to represent the above conceptual
data warehouse models (\emph{star schema} and \emph{snowflake
schema}) with XML. More precisely, we use \emph{XML Schemas} to
define the structure of a data warehouse. To formulate a \emph{star
schema} of a data warehouse in XML, we define the notion of an
\emph{XML star schema} as follows.

\vspace{0,1cm} \emph{Definition: XML star schema}. Let
$(F,\mathcal{D})$ be a star schema, where $F$ is a set of facts
having $m$ measure attributes $\{F.M_{q}, 1 \leq q \leq m\}$ and
$\mathcal{D}=\{D_{s}, 1 \leq s \leq r\}$ is a set of $r$ independent
dimension where each $D_{s}$ contains a set of $n_{s}$ attributes
$\{D_{s}.A_{i}, 1 \leq i \leq n_{s}\}$. The XML star schema of
$(F,\mathcal{D})$ is an XML Schema where: (1) $F$ defines the XML
root element in the XML Schema;   (2) $\forall q \in \{1,\dots,
m\}$, $F.M_{q}$ defines an XML attribute included in the XML root
element;  (3) $\forall s \in \{1,\dots, r\}$, $D_{s}$ defines as
many XML subelements of the XML root element as the number of times
it is linked to the set of facts $F$; (4) $\forall s \in \{1,\dots,
r\}$ and $\forall i \in \{1,\dots, n_{s}\}$, $D_{s}.A_{i}$ defines
an XML attribute included in the XML element $D_{s}$.

Knowing that the XML formalism allows to embed multi-level
subelements in one XML tag, we use this property to represent XML
hierarchies of dimensions. Let $H=\{D_{1}, \dots, D_{t}, \dots,
D_{l}\}$ be a dimension hierarchy. We can represent this hierarchy
by writing $D_{1}$ as an XML element and $\forall t \in \{2,\dots,
l\}$, $D_{t}$ is an XML subelement of the XML element $D_{t-1}$. The
attributes of each $D_{t}$ are defined as XML attributes included in
the XML element $D_{t}$. Since a dimension may have some
hierarchies, it is possible to describe everyone by an XML element
with its embedded subelements. Therefore, we can also define the
notion of \emph{XML snowflake schema}, which is the XML equivalent
of a conceptual \emph{snowflake schema}:

\vspace{0,1cm} \emph{Definition: XML snow flake schema}.  Let
$(F,\mathcal{H})$ be a star schema, where $F$ is a set of facts
having $m$ measure attributes $\{F.M_{q}, 1 \leq q \leq m\}$ and
$\mathcal{H}=\{H_{s}, 1 \leq s \leq r\}$ is a set of $r$ independent
hierarchies. The XML snowflake schema of $(F,\mathcal{H})$ is an XML
Schema where: (1) $F$ defines the XML root element in the XML
Schema; (2) $\forall q \in \{1,\dots, m\}$, $F.M_{q}$ defines an XML
attribute included in the the XML root element; (3) $\forall s \in
\{1,\dots, r\}$, $H_{s}$ defines as many XML dimension hierarchies,
like subelements of the XML root element, as the number of times it
is linked to the set of facts $F$.

\begin{figure*}
\caption{Example of  XML fact}
\label{xml_example}  \vspace{-6,5cm}
\epsfxsize=10cm \centerline{\epsffile[0 0 250 350]{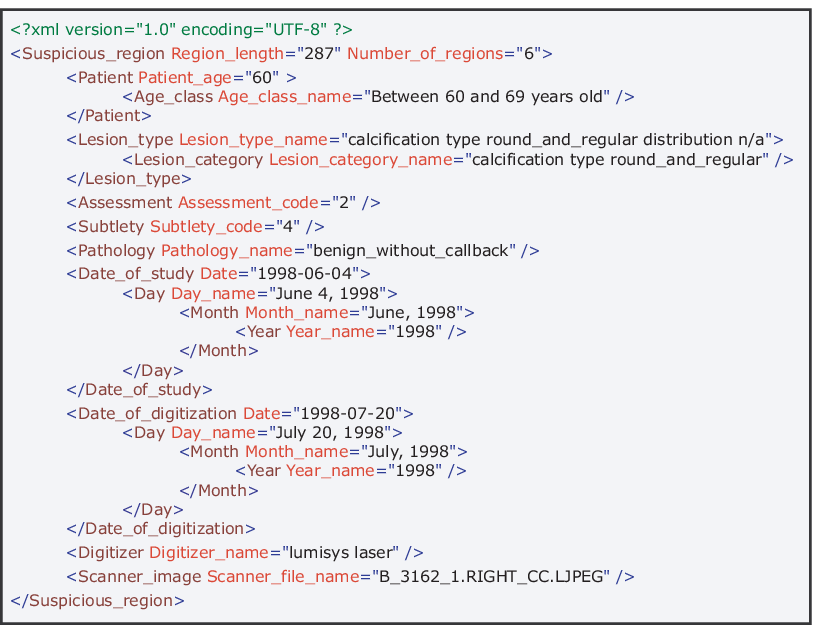}}
\end{figure*}

Based on the properties of the XML formalism, \emph{XML Schemas}
enable to write a logical model of a data warehouse
from its conceptual model. Our approach does not only use
the XML formalism to design data warehouses (or data cubes),
but also feeds them with data. We use XML documents to
support information related to the designed facts.
As an XML document supports values of elements and
attributes, we assume that it contains information
about a single OLAP fact. We say that an XML document
supports an \emph{XML fact} when it is valid against
an \emph{XML star schema} or an \emph{XML snowflake schema}
representing the logical model of a warehouse.
Figure~\ref{xml_example} shows an example of \emph{XML fact}.

\vspace{0,1cm} \emph{Building XML Cubes}. The comparison of
\emph{attribute trees} is realized by fusion operations according to
\emph{pruning} and \emph{grafting} adapted functions
[\ref{boussaid06}]. In some cases, when an input XML document does
not contain enough information required by the analysis objectives,
the fusion provides a poor output XML document, which represents an
OLAP fact with missing data. It is naturally useless to feed the
warehouse with such a document. In order to check wether an input
XML document contains enough information to feed the warehouse or
not, we introduce the \emph{minimal XML document content}. The
\emph{minimal XML document content} is an information threshold
entirely defined by users when submitting the MCM to express
analysis objectives. At this stage, a user can declare for each
measure, dimension, and dimension hierarchy wether it is mandatory
or optional according to his objectives and to the information he
needs to see in the final \emph{XML Cube}. The \emph{minimal XML
document content} corresponds to the \emph{attribute tree}
associated to mandatory elements declared by the user when
submitting the data cube model. It is naturally not possible to
decide with an automatic process which element in a future analysis
context may be optional or not. It is entirely up to the user to
define the \emph{minimal XML document content}. Nevertheless, by
default, we suppose that all measures and dimensions attributes of a
submitted data cube model are mandatory in the final \emph{XML
Cube}. Moreover, we require that not all measures can be optional
elements in the data cube. Indeed, in an analysis context, OLAP
facts without a measure could not be exploited by OLAP operators
such as aggregation. Hence, users are not allowed to set all the
measures to optional elements. At least one measure in the submitted
data cube model must be mandatory.

At the fusion step, the \emph{attribute tree} of an input XML
document is checked. If it contains all the mandatory elements
required by the user, it is merged with the \emph{attribute tree} of
the data cube model. Otherwise, it is rejected, the fusion process
is canceled, and therefore no output document is created. To
validate our approach, we have implemented it as a Java
application~[\ref{boussaid06}].

\subsection{Perspectives}

 A lot of issues need to be addressed in our \emph{X-Warehousing} approach.
 The first perspective is a performance study of OLAP queries in order
 to achieve analysis on XML documents as provided in  \emph{XML Cubes}.
 We should also deal with experimental tests on the reliability of the developed application.
 This includes studies on complexity and time processing of loading input XML
 documents, building attribute trees, merging attribute trees and creating output
 XML documents. Second, we should solve the problem of updating the \emph{XML Cube} when
 the reference MCM is modified in order to change analysis objectives.

 We  also carried out different scenarios concerning different
 XML representations of the physical model in order to study
 the behavior of aggregation queries on the \emph{XML cube}.
 The obtained results seem to highlight the problem of performances
 in \emph{XML cubes}. In the case of certain configurations,
 the results show  that the response time  of the roll-up aggregation
 is prohibitory, but that the physical model is scalable.
 For other configurations, the response time
 of aggregation seems more acceptable but linear. Therefore, the
 concerned configurations are not truly scalable.

 These solutions have the merit to show that the conception
 of \emph{XML cubes} is not trivial. Many problems must be solved.
 The generation of a new aggregate fact must preserve the same
 structure that the original one.
 The detailed facts derived from an aggregate fact must also
 respect the same XML grammar. This latter is definitely defined
 by the cube's  \emph{XML Schema}. When the roll-up or drill down
 operations are achieved, it is necessary to keep the trace of the
 OLAP facts' changes when they are transformed from a granularity level to another into
 the XML cube. This can be achieved through the indices mechanism already used in the
 classical databases.

 Some research already exists that uses the concept of the referential integrity
 (ID/IDREF) to tackle the problem of the response times~[\ref{pokorny02}].
 The idea consists in working with  views of the XML documents that significantly reduce the
 facts by deleting non-useful information for given queries.  The definition of the XML
 documents' views, the fragments of XML views or XML documents have been proposed in
 different articles~[\ref{baril03}].  Combining these mechanisms of indices and views
  would be the solution for better performances in  \emph{XML cubes}.
  The choice of a configuration for the physical model of an XML cube is an open problem.


\section{Complex data analysis}
\label{sec:ComplexDataAnalysis}

After having presented the problems related to warehousing complex
data and the solutions we propose to handle them, we address in this
section the issue of complex data analysis.

The possible approaches to analyze complex data include data mining
and OLAP analysis. Indeed, the integration of complex data into an
ODS under the form of XML documents enables us to consider several
ways to analyze them. The first way consists in exploring  XML
documents directly with data mining techniques. The second way
consists in using OLAP analyses to aggregate complex data and to
explore them in a multidimensional way. However, classical OLAP
tools are ill-adapted to deal with complex data. It seems that OLAP
facts representing complex data need appropriate tools and new ways
of aggregation to be analyzed.\\

\subsection{XML structure mining}

\emph{Context and issues.} The success met by XML is primarily due
to its flexibility and its capacity to describe all kinds of data.
Consequently, it becomes essential to set up suitable techniques to
extract and to exploit information contained in XML documents.
Mining the XML documents concerns the traditional mining techniques,
in particular classification and clustering. There are two main
approaches in XML mining. On one hand, XML content mining  applies
mining methods onto  document contents. On the other hand, XML
structure mining takes interest in information extraction from the
structure of XML documents [\ref{garofalakis99}]. Content mining is
usually based on text mining techniques. Some authors also took
interest in association rules extraction from the content of XML
documents tags [\ref{braga02}]. Nevertheless, this work took very
little  the hierarchical aspect that exists between  tags into
account. Mining the structure of XML documents (intra and
inter-document structure) relates to information contained in the
hierarchical organization of tags [\ref{nayak02}]. The advantage of
this approach is that it  takes the hierarchical structure into
account. In this context, we finally quote some work related to the
extraction of a DTD from a set of XML documents whose structure is
similar [\ref{moh00}].

We are also interested in the extraction of knowledge from the structure
of XML documents. In particular, we wish to release the existing
bonds between tags of a set of homogeneously structured documents.
Among the existing mining methods, association rules extraction
appears the best-adapted in this case. Indeed, this technique proved its
great effectiveness in the discovery of interesting relationships among
a large amount of data.

Association rules extraction from the structure of XML documents
poses specific problems, mainly due to the hierarchical organization
of tags in XML documents. Firstly, XML documents  are not directly
usable for association rules extraction. Indeed, most frequent
itemset search and association rule extraction algorithms are
normally intended to be used within relational databases. They reach
the items  they need for the constitution of  frequent itemsets
thanks to query languages such as SQL (Structured Query Language).
This type of use is not possible in the case of XML documents tags.
To be able to use frequent itemset search  and association rule
extraction algorithms, it is necessary to extract tags from
documents and to structure these data. Thus, it is essential to
carry out a preprocessing step in order to preformat and retrieve
data.

Secondly, XML document tags are hierarchically organized, unlike
tuples in a relational database. Thus, it is necessary to develop a
strategy to manage and to take into account the tags' hierarchical
structure during association rule extraction. Thirdly, in a
well-formed XML document (a document that respects the XML syntax),
the same tag may appear in various places in the hierarchy, but it
always represents the same information.  These tags must be managed
specifically.\\

\begin{figure*}
\caption{Structure mining method in  XML Documents}
\label{mining-xml}  \vspace{-3,5cm} \epsfxsize=8cm \epsffile[0 0 500
600]{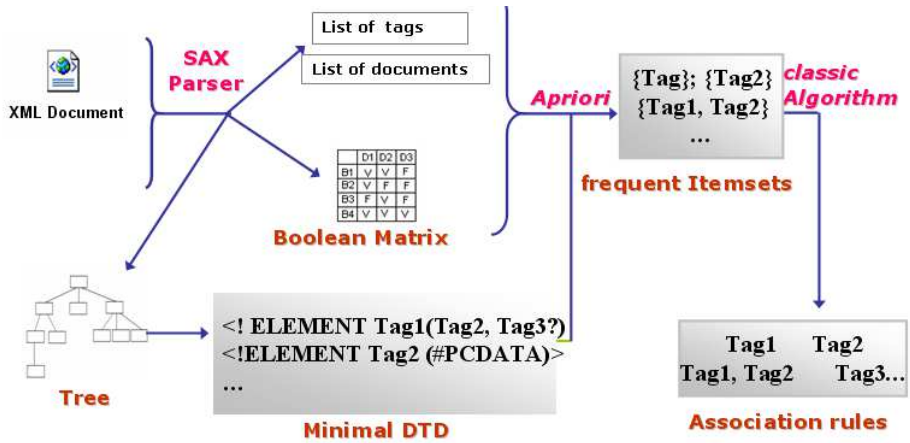}
\end{figure*}

\emph{Proposed solution.} To stage these problems, we carry out an
effective preformating of XML documents and we create a minimal DTD
representing the hierarchy of all the tags met [\ref{duffoux04}].
Moreover, we show that this preformating and the minimal DTD make
them exploitable by traditional extraction algorithms, and set up an
adequate structure for hierarchy management. We obtain a restricted
set of relevant rules while improving processing time. We apply an
adapted version of the \emph{Apriori} algorithm [\ref{agrawal93}]
for frequent itemset search. Lastly, association rules are extracted
and results are presented in XML documents. (Figure~\ref{mining-xml}.\\

In order to test our approach, we used two sets of XML documents: Medline
(the American dictionary of medicine) and a set of French scientific articles.
Then, we applied the adapted traditional algorithm for association
rule extraction. We obtain a set of association rules  we score
thanks to the \emph{discriminating probabilistic indicator} (IPD)
quality measure. We achieve a clear improvement in terms of quality
and processing time for the rules we obtain, compared to a system
that does not use a minimal DTD [\ref{duffoux04}].

Our work can be useful for the creation of a management platform for
XML documents (repairing and creating of a minimal DTD for documents
that do not bear the same schema). This approach also constitutes a
first stage of a broader step in XML mining. The association of
structure and content mining should enable to widen and improve the
XML content mining techniques, in particular for tags that are
bounded. Lastly, in a step of complex data representation, this
mining method can be relevant.\\

\emph{Perspectives.} This approach highlights the interest of mining
the XML document. An XML document encloses more information than an
common text. Its structure supports some relevant information. We
intend to resort to the structure mining as a preliminary task to
enhance the content mining. These both mining tasks constitute  the
efficient XML mining that is different from the traditional text
mining.

The structure mining may be perceived  as an interesting way to
discover the relevance of some tags, which they may be selected as
measures or dimensions in the multidimensional modeling task
assisted by data mining techniques [\ref{boussaid06jgo}]. In order
to achieve this assisted modeling, we can exploit the associations
among some tags discovered by our structure mining approach.\\

\subsection{A data mining-based OLAP operator for complex data}

\emph{Context and issues.} OLAP is a powerful mean of exploring and
extracting pertinent information from data through multidimensional
analysis. In this context, data are organized in multidimensional
views, commonly called data cubes. However, classical OLAP tools are
not always able to deal with complex data. For example, when
processing images, sounds, videos, texts or even XML documents,
aggregating information with the classical OLAP does not make any
sense. Indeed, we are not able to compute a traditional aggregation
operation, such as sum or average, over such data. However, when
users analyze complex data, they need more expressive aggregates
than those created from elementary computation of additive measures.
We think that OLAP facts representing complex objects need
appropriate tools and new aggregation means since we wish to analyze
them.

Furthermore, a data cube structure can provide a suitable context
for applying data mining methods. More generally, the association of
OLAP and data mining allows elaborated analysis tasks exceeding the
simple exploration of a data cube. Our idea is to take advantage
from OLAP, as well as data mining techniques and to integrate them
into the same analysis framework in order to analyze complex
objects. Despite the fact that both OLAP and data mining have long
been considered two separate fields, several recent studies proved
the capability of their association to provide interesting analysis
process [\ref{imielinski96}]. The major difficulty when combining
OLAP and data mining is that traditional data mining algorithms are
mostly designed for tabular datasets organized in
individual-variable form [\ref{fayyad96}]. Therefore,
multidimensional data are not suited to these algorithms.
Nevertheless, a lot of previous research motivated and proved an
interest for coupling OLAP with data mining methods in order to
extend OLAP analysis capabilities [\ref{goil98}, \ref{han98},
\ref{sarawagi98}]. In addition, we propose another contribution to
this field by developing a new type of online aggregation for
complex data. It is a new OLAP Operator for Aggregation by
Clustering (OpAC) [\ref{benmessaoud04}].\\

\emph{Proposed solution.} To aggregate information about complex
data, we must often gather similar facts into a single group and
separate dissimilar facts into different groups. In this case, it is
necessary to consider an aggregation by computing both descriptors
and measures. Instead of grouping facts only by computing their
measures, we also take their descriptors into account to obtain
aggregates expressing semantic similarities. Our new operator for
aggregating complex data (OpAC) combines OLAP with an automatic
clustering technique. We use the Agglomerative Hierarchical
Clustering (AHC) as an aggregation strategy for complex data. Our
operator enables significant aggregates of facts expressing semantic
similarities. More generally, the aggregates provided by OpAC give
interesting knowledge about the analyzed domain. OpAC is adapted for
all types of data, since it deals with data cubes modeled in XML
[\ref{benmessaoud06}].

 Furthermore, we also propose some evaluation criteria that support
the results of our operator. These criteria aim at assisting the
user and helping him/her in choosing the best partition of aggregates
that will fit well with his/her analysis requirements. We also
developed a Web application for our operator. We provided
performance experiments and drove a case study on XML documents
dealing with the breast cancer research domain [\ref{benmessaoud06}] .

The main idea of  OpAC is to
exploit the cube's facts describing complex objects  to provide over them a
more significant aggregation. In order to do so, we use a clustering method and
automatically highlight aggregates that are semantically richer than those
provided by the current OLAP operators. Hence,  the clustering method provides
a new OLAP aggregation concept. This aggregation provides hierarchical groups
of objects resuming information and enables navigating through levels of these
groups. Existing OLAP tools, such as the Slicing operator, can create new
restricted aggregates in a cube dimension, too, but these tools always need a
handmade assistance, whereas our operator is based on a clustering algorithm
that automatically provides  relevant aggregates. Furthermore, with classical
OLAP tools, aggregates are created in an intuitive way in order to compare some
measure values, whereas OpAC creates significant aggregates that express deep
relations with the cube's measures. Thus, the construction of such aggregates
is interesting to establish a more elaborate on-line analysis context.
According to the above objectives, we choose the AHC as an aggregation method.
Our choice is motivated by the fact that the hierarchical aspect constitutes a
relevant analogy between AHC results and the hierarchical structures of
dimensions. The objectives and the results expected for OpAC match perfectly
with the AHC strategy. Furthermore, the AHC adopts an agglomerative strategy
that starts by the finest partition where each individual is considered a
cluster. Therefore, OpAC results include the finest attributes of a dimension.
But, like almost all unsupervised mining methods, the main defect of the AHC is
that it does not give an evaluation of its results, i.e. the partitions of
clusters. It is quite tedious to choose the best partition, and it is more
difficult when we deal with a great number of individuals. In the case of OpAC
operator, we propose to use many criteria to help users to select  the best
partition of aggregates. According to the data mining literature, we can cause
the intra and inter-clusters inertias or the Ward's method. In addition, we
also propose a new criterion based on the cluster separability
[\ref{benmessaoud06}].

The AHC is compatible with the exploratory aspect
of OLAP. Its results can also be reused by classical OLAP operators.
In fact, the AHC provides several hierarchical partitions. By moving
from a partition level to a higher one, two aggregates are joined
together. Conversely, by moving from a partition level to a lower
one, an aggregate is divided into two new ones. These operations are
strongly similar to the classical Roll-up and Drill-down operators.
AHC is a well suited clustering method to summarize information into
OLAP aggregates from complex facts.

\emph{Perspectives.} This work has proved the interest of
associating OLAP and data mining in order to enhance on-line
analysis power. We believe that, in the future, this association
will provide a new generation of efficient  OLAP operators
well-suited to complex data analyses. To extend the on-line analysis
toward the new capabilities such as explanation and prediction is an
exciting challenge that we plan to achieve. In order to aim this
objective, we intend to develop  a new algebra based on the OLAP and
data mining coupling to define some new on-line mining operators to
deeply explore the complex data.


\section{Performance issues in XML warehousing}
\label{sec:PerformanceIssues}

\subsection{Context and issues}
\label{pii}

While we advocated in the previous sections that XML data warehouses
form an interesting basis for decision-support applications exploiting complex data,
XML-native database management systems (DBMSs) usually show limited performances
when the volume of data is very large and queries are complex.
This is typically the case in data warehouses, where data are historicized
and analytical queries involve several join and aggregation operations.
Hence, it is crucial to devise means of optimizing the performance of
XML data warehouses. In such a context, indexing and view materialization
are presumably some of the most effective optimization techniques.

Indexes are physical structures that allow direct data access.
They avoid sequential scans and thereby reduce query response time.
Materialized views improve data access time by precomputing
intermediary results. Therefore, end-user queries can be
efficiently processed through data stored in views, and do
not need to access the original data. However, exploiting
either indexes or materialized views requires additional
storage space and entails maintenance overhead when refreshing
the data warehouse. The issue is then to select an appropriate
configuration of indexes and materialized views that minimizes
both query response time and index and view maintenance cost,
given a limited storage space (a NP-hard problem).

Most of the existing XML indexing techniques [\ref{bertino04},
\ref{boulos06}, \ref{bruno02}, \ref{chien02}, \ref{chung02},
\ref{goldman97}, \ref{he04}, \ref{jiang03}, \ref{kaushik02},
\ref{milo99}, \ref{qun03}, \ref{zhang01}] are applicable only
onto XML data that are targeted by simple path expressions.
However, in the context of XML data warehouses, queries are
complex and include several path expressions. Moreover,
these indices operate on one XML document only, whereas
in XML warehouses, data are managed in several XML documents
and analytical queries are performed over these documents.
The Fabric index [\ref{cooper01}] does handle multiple documents,
but it is not adapted to XML data warehouses either, because
it does not take into account the relationships that exist
between XML documents in a warehouse (facts and dimensions).
Fabric is thus not beneficial to decision-support queries.
In consequence, we propose a new index that is specifically
adapted to XML, multidimensional data warehouses [\ref{mahboubi06egc}].
This data structure allows to optimize the access time to several
XML documents by eliminating join costs, while preserving the
information contained in the initial warehouse.

Eventually, the literature about materialized view selection
is abundant in the context of relational data warehouses but,
to the best of our knowledge, no such approach exists in XML
databases and XML data warehouses in particular. Hence,
we proposed an adaptation of a query clustering-based relational
view selection approach [\ref{aouiche06}] to the XML context
[\ref{mahboubi06}]. This approach clusters XQuery queries and
builds candidate XML views that can resolve multiple similar
queries belonging to the same cluster. XML-specific cost models
are used to define the XML views that are pertinent to materialize.

\subsection{Proposed solutions}

\emph{Reference XML data warehouse.} Several authors address
the issue of designing and building XML data warehouses.
They use XML documents to manage or represent the facts
and/or dimensions of the warehouse
[\ref{hummer03}, \ref{park05}, \ref{pokorny02}].
We select XCube [\ref{hummer03}] as a reference data
warehouse model. Since other XML warehouse models from
the literature are relatively similar, this is not a
binding choice. The advantage of XCube is its simple
structure for representing facts and dimensions in a
star schema: one document for dimensions (\emph{Dimensions.xml})
and another one for facts (\emph{Facts.xml}).

\begin{figure*}[hbt]
\caption{\emph{Facts.xml} structure (a) and its corresponding 1-index (b)}
\label{xcube}
\centerline{\epsffile{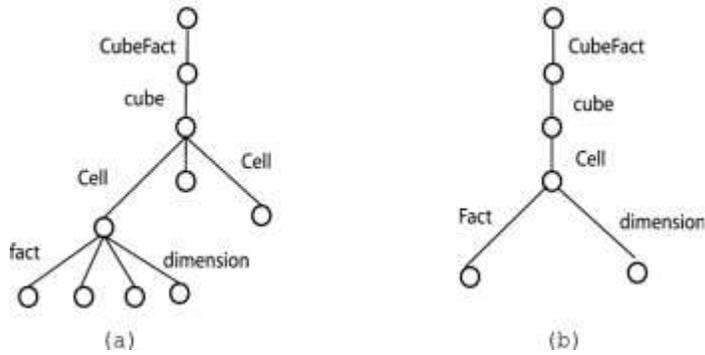}}
\end{figure*}

\emph{XML data warehouse indexing.} Building the indexes
cited in Section~\ref{pii} on an XML warehouse causes a
loss of information in decision-support query resolution.
Indeed, clustering or merging identical labels in the XML
graph causes the disappearance of the relationship between
fact measures and dimensions. We illustrate this problem
in the following example. The \textit{Facts.xml} document
is composed of \textit{Cell} elements. Each cell is
identified by its attributes and one or more measures.
Figure~\ref{xcube} shows the structure of the \textit{Facts.xml}
document and its corresponding 1-index (we selected the 1-index
[\ref{milo99}] as an example). The 1-index represents cells linearly,
i.e., all labels for the same source are represented by only one
label. Hence, recovering a cell characterized by its measures and
its dimension identifiers is impossible.

\begin{figure*}[hbt]
\caption{XML join index structure}
\label{index}
\centerline{\epsffile{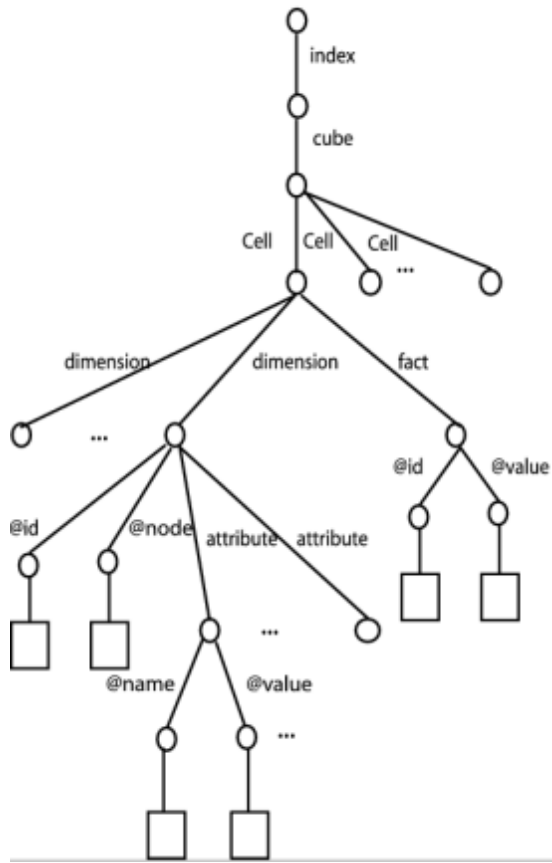}}
\end{figure*}

An index should be able to preserve the relationships between dimensions and fact measures.
Thus, our index' structure is similar to that of the \textit{Facts.xml} document, except for
the \textit{attribute} element. As usual with XML indexes, our index structure is stored in
an XML document named \textit{Index.xml} (Figure~\ref{index}). Each \textit{Cell} element
is composed of dimensions and one or more facts. A \textit{Fact} element has two attributes,
$@id$ and $@value$, which respectively represent measure names and values. Each \textit{dimension}
element is composed of two attributes: $@id$, which stores the dimension name, and $@node$,
which stores the value of the dimension identifier. Moreover, the dimension element has
children \textit{attribute} elements. These
elements are used to store the names and values of the attributes from each dimension.
They are obtained from the \textit{Dimensions.xml} document. An \textit{attribute} element
is composed of two attributes, $@name$ and $@value$, which respectively store the name and
value of each attribute.

Data migration from \textit{Dimensions.xml} and \textit{Facts.xml}
to the index structure helps in storing facts, dimensions and their
attributes in the same cell. This feature wholly eliminates join
operations, since all the information that is necessary for a join
operation is stored in the same cell. Finally, queries need to be
rewritten to exploit our index. The rewriting process consists in
preserving the selection expressions and the aggregation operations.

\emph{XML materialized view selection.} The first step in our view
selection strategy is to build, from a workload of representative
queries, a clustering context. We extract from the queries representative
attributes, i.e., attributes that are present in the selection predicates
and grouping clauses. Then, we store the relationships between the query
workload and the extracted attributes in a ``query-attribute" matrix.
The matrix' lines are the queries and the columns are the extracted attributes.
The general term $q_{ji}$ of this matrix is set to one if extracted attribute
$a_{j}$ is present in query $q_{i}$, and to zero otherwise. This matrix
represents our clustering context.

Then, since it is hard to search for all the syntactically relevant views
(candidate views) because the search space is very large, we cluster
together similar queries. Similar queries are the one having a close
binary representation in the query-attribute matrix. Two similar queries
can be resolved by using only one materialized view. We define similarity
and dissimilarity measures that ensure that queries within a same cluster
are strongly related to each others, whereas queries from different clusters
are significantly different. The number of candidate views we obtain is
generally as high as the input workload is large. Thus, it is not feasible
to materialize all the proposed views because of storage space constraints.
To circumvent this limitation, we devised cost models and objective
functions that exploit them, and help in selecting only the most pertinent
materialized views.

Finally, our view selection algorithm is based on a greedy search within
the candidate view set $V$, with respect to an objective function $F$.
In the first iteration, the values of $F$ are computed for each view
within $V$. The view $v_{max}$ that minimizes $F$, if it exists
($F_{/S}(v_{max}) > 0$), is then added to $S$. The values of $F$
are then computed for each remaining view in $V - S$, since they
depend on the selected views present in $S$. This helps in taking
into account the interactions that probably exist between the views.
We repeat these iterations until there is no improvement
($F_{/S}(v) \leq 0$), all the views have been selected
$(V - S = \emptyset)$ or storage space is full.

\subsection{Perspectives}

To validate our proposals, we performed many experiments with
XML-native and relational, XML-compatible DBMSs. Our tests show
that using either the index structure we propose or the materialized
views our strategy helps in building significantly improves response
time for typical analytical queries expressed in XQuery. Gains
in performance indeed range from a factor 8,000 to 25,000, depending
on the host system.
Furthermore, our tests also demonstrate that, properly indexed,
XML-native DBMSs can compete with, and even best relational DBMSs
in terms of performance when XML documents are bulky. This is
because relational DBMS engines combine XQuery to SQL and must
convert the result from relations to XML. Native-XML DBMSs,
on the other hand, preserve the hierarchical structure of XML
data, which allows path scans to be efficiently processed by
 XQuery engines.

This research opens three broad axes of perspectives. First,
we have to complement our experiments with other tests, using
presumably other systems and data warehouse configurations.
Our aim is to assert the gain in performance \emph{vs.}
the overhead for generating and refreshing indexes and
materialized views, in each configuration.
Second, it is widely acknowledged that indexes and materialized
views are mutually beneficial to each other. We have designed
a method for simultaneously selecting indexes and materialized
views in the relational context, which we aim at adapting to
the XML context.
Finally, our performance optimization strategies could be
better integrated in a host XML-native DBMS. This would
certainly help in developing an incremental strategy for
the maintenance of indexes and materialized views.
Our selection strategy is indeed static, currently.
Studies dealing with incremental data mining may be
exploited to make it dynamic, so that a configuration
of indexes and materialized views can be updated instead
of being rebuilt from scratch. Moreover, the mechanism
for rewriting queries would also be more efficient if
it was part of the system.


\section{Conclusion}
\label{sec:Conclusion}

Nowadays, data from the Web are complex. To handle this kind of data
in a decision-support context, we propose a full, generic data
warehousing and on-line analysis process. In this paper, we
identified some problems related to both these  processes and we
suggested different solutions.

First of all, we exposed an approach to integrate complex data. We
presented a generic UML model that allows to model not only
low-level but also semantic information concerning the complex data
to be analyzed. We integrated complex data as XML documents into an
ODS as a first step in complex data warehousing. This complex data
integration approach is based on both the data warehouse technology
and multi-agent systems. Our integration approach indeed
necessitates several tasks that may be assimilated to services
offered by well-defined agents in a system intended to achieve such
an integration. We developed the SMAIDoC system, which allows this
integration. It is based upon a flexible and progressive
architecture on which we can add, remove or modify services, and
even create new agents.

Secondly, we proposed a methodology entirely based on the XML
formalism to warehouse complex data. Our X-Warehousing approach does
not simply populate a repository with XML documents, but also
expresses an interesting abstraction level by preparing XML
documents to future analyses. In fact, it consists in validating
documents against an XML Schema, which models a data warehouse. We
defined a general XML formalization for star and snowflake schemas.
We also exploited the concept of attribute trees to help in the
creation and the warehousing of homogeneous XML documents, by
merging initial XML sources with a reference multidimensional model.
Constraints on the created XML documents may be required and
expressed by users. To validate our X-Warehousing approach, we
implemented a Java application, which takes as input a reference
multidimensional model and XML documents, and provides logical and
physical models for an XML cube composed of homogeneous XML
documents.

These proposed solutions constitute the first axis related to the
complex warehousing process. We also proposed two approaches
regarding the on-line analysis process of complex data. The first
one applies a data mining technique to discover association rules
among the tags of XML documents. Structure mining is indeed a
relevant preliminary task for content mining over XML documents. Our
second proposal carries on a new on-line analysis context for
complex data. Our approach is based on coupling OLAP with data
mining. The combination of the two fields can be a solution to
capitalize their respective benefits. We have created OpAC, a new
OLAP aggregation operator based on an automatic clustering method.
Unlike classical OLAP operators, OpAC enables precise analyses and
provides semantic aggregates of complex objects.

Finally, the implementation of our X-Warehousing method highlighted
performance problems when storing warehouses in XML-native DBMSs.
Hence, we proposed a new join index that is specifically adapted to
XML, multidimensional data warehouses. This data structure allows to
optimize the access time to several XML documents by eliminating
join costs, while preserving the information contained in the
initial warehouse. We also presented a view selection algorithm,
which we combine with the index structure to improve the overall
efficiency of our strategy. The experiments we performed show a
significant improvement in response time for analytical queries.
\\

All the solutions we have presented in this paper are a modest
contribution to the problem of complex data warehousing and on-line
analysis. Although we mentioned various perspectives for our work at
the end of each of this paper's sections, a lot of scientific issues
are still not solved.


\section*{Acknowledgements}

The authors would like to thank Prof. Robert Wrembel and Prof. Jaroslav Pokorn\'{y}, the editors, for inviting them to publish an article in this special issue.


\NUMBIB

\end{document}